# Using artificial intelligence for data reduction in mechanical engineering

*L. Mdlazi[1], C.J. Stander[1], P.S. Heyns[1], T. Marwala[2]*

[1] Dynamic Systems Group
Department of Mechanical and Aeronautical Engineering, University of Pretoria
Pretoria, 0002, South Africa
E-mail: Lungile.Mdlazi@eskom.co.za

[2] School of Electrical and Information Engineering,
University of the Witwatersrand, Private Bag 3, Wits, 2050, South Africa
E-mail: t.marwala@ee.wits.ac.za

**Abstract**
In this paper artificial neural networks and support vector machines are used to reduce the amount of vibration data that is required to estimate the Time Domain Average of a gear vibration signal. Two models for estimating the time domain average of a gear vibration signal are proposed. The models are tested on data from an accelerated gear life test rig. Experimental results indicate that the required data for calculating the Time Domain Average of a gear vibration signal can be reduced by up to 75% when the proposed models are implemented.

## 1. Introduction

Calculating the Time Domain Average (TDA) of a gear vibration signal by direct averaging using digital computers requires large amounts of data [1-7]. This requirement makes it difficult to develop online gearbox condition monitoring systems that utilize time domain averaging calculated by direct averaging to enhance diagnostic capability. This study presents a novel approach to estimating the TDA of a gear vibration signal, using less data than would be used when calculating the TDA by direct averaging.

Artificial Neural Networks (ANNs) and Support Vector Machines (SVMs) are used for estimating the TDA of a gear vibration signal. Two models are presented. The input data comprises rotation-synchronized gear vibration signals and the output is the TDA of the gear vibration signal. When Model 1 is used, the results indicate that the amount of gear vibration data required for calculating the TDA is reduced to 25 percent of the amount of data required when calculating the TDA by direct averaging. When Model 2 is used, the amount of data to be stored in the data acquisition system is reduced to less than 20 percent of the data that would be stored when calculating the TDA by direct averaging. The ANNs that are implemented are Multi-Layer Perceptrons (MLPs) and Radial Basis Functions (RBFs) [8]. Two parameters were selected to verify whether the TDA estimated by the models retains the original diagnostic capability of the TDA. These parameters are the kurtosis for impulses and the peak value for overall vibration. The computational time is also compared to determine the suitability of the proposed models in real-time analysis.

## 2. ANNs and SVMs

ANNs and SVMs may be viewed as parameterized non-linear mapping of input data to the output data. Learning algorithms are viewed as methods for finding parameter values that look probable in the light of the data. The learning process takes place by training the ANNs or SVMs through supervised learning. Supervised learning is the case where the input data set and the output data set are both known, and ANNs or SVMs are used to approximate the functional mapping between the two data sets.

### 2.1 Multi-layer Peceptron

A two-layered MLP architecture was used. This selection was made because of the universal approximation theorem, which states that a two-layered architecture is adequate for the MLP. The MLP provides a distributed representation with respect to the input space due to the cross-coupling between hidden units. The output of a two-layer perceptron can be expressed as the following equation:

$$y_k = f_{outer}\left(\sum_{j=1}^{M} w_{kj}^{(2)} f_{inner}\left(\sum w_{ji}^{(1)} x_i + w_{j0}^{(1)}\right) + w_{k0}^{(2)}\right) \quad (1)$$

where $f_{outer}$ and $f_{inner}$ are activation functions, $w_{ji}^{(1)}$ denotes a weight in the first layer, going from input $i$ to hidden unit $j$, $w_{k0}^{(2)}$ is the bias for the hidden unit $k$ and $w_{kj}^{(2)}$ denotes a weight in the second layer. The parameter $f_{inner}$ was selected as hyperbolic tangent function "tanh" and $f_{outer}$ was selected as a linear function. The hyperbolic tangent function maps the interval [-∞,∞] onto the interval [-1,1] and the linear activation function maps the interval [-∞,∞] onto the interval [-∞,∞]. The maximum-likelihood approach was used for training the MLP network. The sum-of-squares-of-error and the weight decay regularization was used as cost functions [8, 9]. The weight decay penalizes large weights and ensures that the mapping function is smooth, avoiding an over-fitted mapping between the input data and the output data [9]. In this study, a regularization coefficient of 1.5 was found most suitable. The weights $w_i$ and biases in the hidden layers were varied using Scaled Conjugate Gradient (SCG) optimization until the cost function was minimized [10]. It was determined empirically that a two-layer MLP network with five hidden units was best suited to this application.

### 2.2 Radial basis functions

The RBF network approximates functions by a combination of radial basis functions and a linear output layer. The RBF neural networks provide a smooth interpolating function for which the number of basis functions is determined by the complexity of the mapping to be represented, rather than by the data set. The RBF neural network mapping is given by

$$y_k(\mathbf{x}) = \sum_{j=1}^{M} \varpi_{kj} \phi_j(\mathbf{x}) + \varpi_{k0} \quad (2)$$

where $\varpi_{k0}$ are the biases, $\varpi_{kj}$ are the output layer weights, $\mathbf{x}$ is the d-dimensional input vector and $\phi_j(\cdot)$ is the j$^{th}$ basis function. The thin plate-spline basis function was used in this study [8]. The radial basis function is trained in two stages. In the first stage the input data set $\mathbf{x}^n$ alone is used to determine the basis function parameters. After the first training stage, the basis functions are kept fixed and the second layer of weights is determined in the second training phase. Since the basis functions are considered fixed, the network is equivalent to a single-layer network that can be optimized by minimizing a suitable error function. The sum-of-square error function is also used to train RBFs. The error function is a quadratic function of the weights and its minimum can therefore be found in terms of the solution of a set of linear equations. For regression the basis function parameters are found by treating the basis function centers and widths, along with the second-layer weights, as adaptive parameters to be determined by minimizing the error function. In this study it was determined empirically that a RBF network with five basis functions was most suitable.

### 2.3 Support vector machines

SVMs were developed by Vapnik [9] and have gained much popularity in recent years. The SVM formulation embodies the Structural Risk Minimization (SRM) principle, which has been shown to be superior to the traditional Empirical Risk Minimization (ERM) principle employed by conventional neural networks [9, 11, 12]. SRM minimizes an upper limit on the expected risk, as opposed to the ERM that minimizes the error on the training data. It is this difference that gives SVMs a greater ability to generalize. When SVMs are applied to regression problems, loss functions that include a distance measure are used. The ε-insensitive loss function [9] was selected for this study. The ε-insensitive loss function is defined by:

$$L_\varepsilon(y) = \begin{cases} 0 & \text{for } |f(\mathrm{x}) - y| \\ |f(\mathrm{x}) - y| - \varepsilon & \text{Otherwise} \end{cases}. \quad (3)$$

In non-linear regression, non-linear mapping is used to map the data to a higher dimensional feature space where linear regression is performed. The kernel approach is employed to address the curse of dimensionality. The non-linear support vector regression solution, using the ε-insensitive loss function, is given by:

$$\max_{\alpha,\alpha^*}(\alpha,\alpha^*) = \max_{\alpha\alpha^*} \sum_{i=1}^{l}\sum_{j=1}^{j} \alpha^*(y_i - \in) - \alpha_i(y_i - \in) \quad (4)$$

$$-\frac{1}{2}\sum_{i=1}^{l}\sum_{j=1}^{j}(\alpha_i^* - \alpha_i)(\alpha_j^* - \alpha_j)K(x_i, x_j)$$

with constraints,

$$0 \leq \alpha_i, \alpha_i^* \leq C, \quad i = 1, K, l \quad (5)$$

$$\sum_{il=1}^{l}(\alpha_i - \alpha_i^*) = 0$$

Solving Equation (3) with the constraints in Equation (5) determines the Lagrange multipliers, $\alpha$ and $\alpha^*$ and the regression function is given by

$$f(x) = \sum_{i=1}^{i}(\overline{\alpha}_i - \overline{\alpha}_i^*)K(x_i, x) + \overline{b} \quad (6)$$

where

$$\langle \overline{w}, x \rangle = \sum_{SVs}(\alpha_i - \alpha_i^*)K(x_i, x_j) \quad (7)$$

$$\overline{b} = -\frac{1}{2}\sum_{i=1}(\alpha_i - \alpha_i^*)(K(x_i, x_r) + K(x_i, x_s)).$$

Different kernels were investigated for mapping the data to a higher dimensional feature space where linear regression was performed. The exponential radial basis function kernel [11] with an order of 10 was found most suitable for this application.

## 3. Proposed models

Two different models are proposed. Model 1 maps the input space to the target using simple feed-forward network configuration. The size of the input space is systematically reduced to find the optimal number of input vectors that can be used to estimate the target vector correctly. When the ANNs and SVMs are properly trained, Model 1 is capable of mapping the input space to the target, using less data than would otherwise be used when calculating the TDA by direct averaging. It was determined empirically that 40 rotation-synchronized gear vibration signals were suitable for predicting the TDA with Model 1. Consequently, the amount of data required for calculating the TDA is reduced to 25 percent when using 40 rotation-synchronized gear vibration signals, since 160 rotation-synchronized gear vibration signals were used in calculating the TDA by direct averaging. Figure 2 shows a schematic diagram of the proposed methodology.

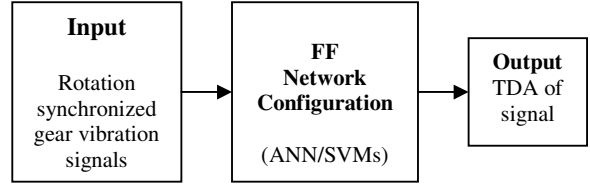

Figure 1. Schematic of proposed methodology

Model 2 estimates the TDA of the input space in small sequential steps, analogous to taking a running average of the input space. This model consists of a number of feed-forward networks. Instead of using the network to estimate the TDA of the entire input data, Model 2 first sequentially estimates the average of subsections of the input data. The output of the first stage is used as input into the second network that estimates the TDA of the entire input data. The feed-forward networks are trained off-line to reduce computation time. In Model 2, data can be discarded immediately after use. This means that the model does not require large amounts of data to be stored in the data logger, even though all the data should be collected. In this study, 10 rotation-synchronized gear vibration signals were found most suitable for estimating the instantaneous average in the first stage of estimation. As a result, the amount of vibration data that is stored in the data logger was reduced to less than 20 percent of the amount of data that is stored in the data logger when calculating the TDA by direct averaging.

## 4. Estimation results

Model 1 and Model 2 were used for estimating the TDA of the gear vibration signal so that the TDA estimated by the proposed models could be compared with the TDA calculated by direct averaging. The data that was used was from the accelerated gear life test rig [13, 14]. Comparisons were made in the time and the frequency domains. To quantify the accuracy of the TDA estimated by the models, the 'fit' parameter $\eta_{sim}$ [15] defined by

$$\eta_{sim} = 100 \frac{\sum_{k=1}^{N}|e(k)|}{\sum_{k=1}^{N}|y_{desired}(k)|} \quad [\%], \quad (8)$$

was used. In Equation 8 $e(k)$ is the simulation accuracy defined by Equation (9) and $N$ is the number of data points used.

$$e(k) = y_{desired}(k) - y_{achieved}(k) \qquad (9)$$

$y_{desired}$ is the TDA signal calculated by direct averaging and $y_{achieved}$ is the TDA signal estimated by the models. The 'fit' parameter $\eta_{sim}$ gives a single value for each simulation, therefore can be used to compare the performance of the different formulations over the entire gear life. A high value for the 'fit' parameter $\eta_{sim}$ implies a bad fit whereas a low value implies a good fit. Through trial and error it was established that $\eta_{sim} = 40\%$ is a suitable upper cut-off point for simulation accuracy.

For Model 1, 40 rotation-synchronized gear vibration signals from the first test with 1024 points per revolution were used for training the ANNs and 256 points per revolution were used to train SVMs. This resulted in training sets of dimensions 1×40×1024 and 1×40×256 respectively. Fifteen test data sets were measured through the life of the gear and used as validation sets. This resulted in validation data sets of dimensions 15×160×1024 15×160×256 for ANNs and SVMs respectively. For Model 2, the whole data set of 160 rotation-synchronized gear vibration signals from the first test was used for training the ANNs and SVMs. This resulted in a training set of dimensions 1×160×1024 for ANNs and 1×160×256 for SVMs. The rest of the data of dimensions 15×160×1024 and 15×160×256 were used as validation data for ANNs and SVMs respectively.

Figure 2 shows the estimation results obtained when Model 1 with an MLP network was simulated using unseen validation data sets of 40 input signals during the running-in stages of gear life. The dotted line is the TDA estimated by Model 1, and the solid line is the TDA calculated by direct averaging. The first plot in Figure 2 is the time domain representation of the results; the second plot is the frequency domain representation. It is observed that both the time and frequency domain representations are almost exact fits. This shows that Model 1 with MLP networks retains the time and frequency domain properties of the original time domain averaging process when using gear vibration data from the accelerated gear life test rig.

Similar performances were obtained throughout the life of the gear using both models with RBFs and SVMs, even though there were significant changes in the vibration signatures as the condition of the monitored gear deteriorated. The changes in the vibration signatures were due to changes in the meshing stiffness caused by cracks in the gear teeth. The good performance is due to the mapping and generalization capabilities of ANNs and SVMs.

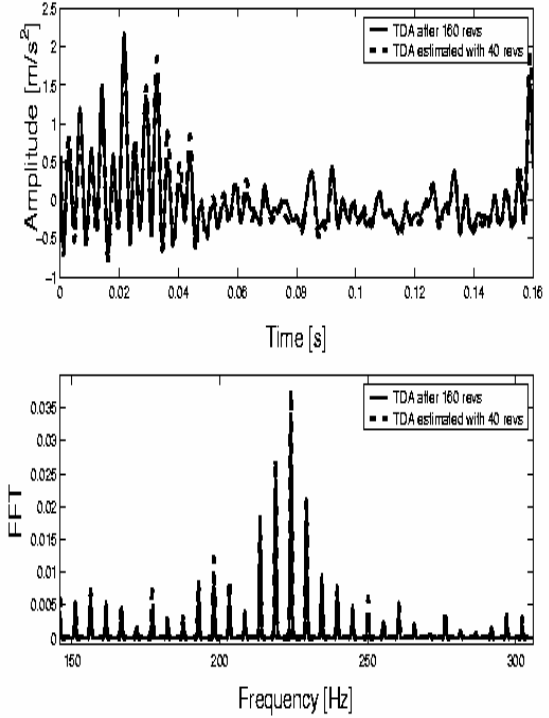

Figure 2. Model 1 with MLP estimation using data from a test conducted under constant load conditions

Figure 3 shows the simulation accuracy $\eta_{sim}$ plotted against the gear life for Model 1. It is observed that Model 1 with RBF network and Model 1 with SVMs give similarly performance. Their performance was slightly better than that of Model 1 with MLP networks. The performances of all three formulations are acceptable because $\eta_{sim}$ is less than the cut-off value for all the formulations.

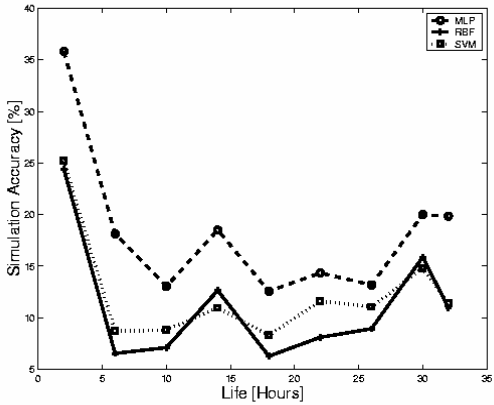

Figure 3. Model 1 Simulation accuracy $\eta_{sim}$ vs. gear life under constant load conditions

In addition to simply considering the goodness of fit, the diagnostic capabilities of the TDA estimated by the models was assessed using the peak value of the vibration $X_{max}$ during a given interval $T$ and the Kurtosis. The peak value is used to monitor the overall magnitude of the vibration to distinguish between acceptable and unacceptable vibration levels and the kurtosis is useful for detecting the presence of an impulse within the vibration signal [16]. The peak value of the vibration $X_{max}$ and the kurtosis of the TDA calculated using direct averaging, were compared to the TDA estimate from the proposed models.

Figure 4 is a plot of $X_{max}$ and kurtosis calculated from the TDA estimated by Model 1, superimposed on the $X_{max}$ and kurtosis calculated from the TDA obtained by using direct averaging. Figure 4 indicates that the kurtosis is an exact fit for all three formulations. This implies that the TDA predicted by Model 1 can be used to monitor the presence of impulses in the measured gear vibration signal. It is also observed that the kurtosis is very high during the early stages of gear life. This is characteristic of the running-in stages of the gear life, during which the vibration signature tends to be random. A similar trend is observed during the wear-out stage in which strong impulses are caused by the reduction in stiffness in cracked or broken gear teeth. Only the peak values obtained from Model 1 with MLP and SVM are close fits and can be used to monitor the amplitude of the overall vibration. Model 1 with RBF achieved an unsatisfactory performance because the RBF network selected in this simulation was not optimal; consequently it did not generalize well to changes in the measured vibration as gear failure progressed. Similar results were obtained using Model 2.

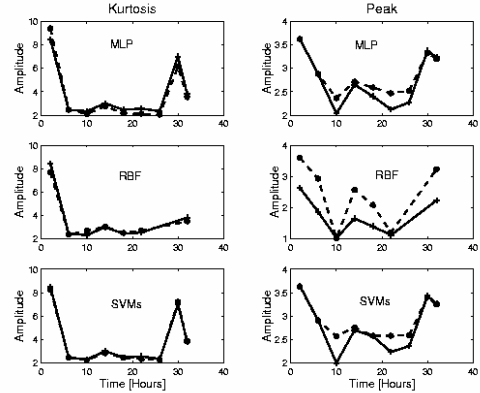

Figure 4. Comparison of kurtosis and peak values for the TDA calculated by direct averaging (+) and the TDA predicted by Model 1 (*) with MLP, RBF and SVMs

To put the proposed models into perspective, their computation time was compared to that of the existing time domain averaging method using a Pentium 4 computer with a 1.60 GHz processor. The computation times are listed in Table 1.

Table 1. Computation time in seconds

| Time parameter [s] | TDA | MLP | RBF | SVM |
|---|---|---|---|---|
| Pre-processing Model 1 | 1.011 | 0.703 | 0.703 | 0.703 |
| Training Model 1 | - | 22.24 | 2.219 | 497.0 |
| Simulating Model 1 | 0.75 | 0.016 | 0.047 | 5.500 |
| Pre-processing Model 2 | 1.011 | 1.011 | 1.011 | 1.011 |
| Training Model 2 | - | 1.14 | 1.015 | 963.8 |
| Simulating Model 2 | - | 0.08 | 0.078 | 83.76 |

It is clear that Model 1 requires less pre-processing time than the TDA calculated by direct averaging. This is because Model 1 uses 25 percent of the vibration data while the original TDA process uses all of the vibration data. The required pre-processing time for Model 2 is equal to that of calculating the TDA by direct averaging because both models use the same amount of vibration data. When Model 1 and Model 2 are used, RBF and MLP give the best performance in terms of simulating time and SVMs give the poorest performance. The models are trained off-line therefore the training time would not influence the performance in real-time applications. When the models are used with SVMs their performance is poor. The poor performance with SVMs is due to the optimization problem. In SVMs optimization is a quadratic problem with 2N variables, where N is the number of data training points. Longer training times are needed because more operations are required in the process. This is much slower than the MLP and RBF neural networks in which only the weights and biases or the basis centers are obtained by minimizing the error functions.

## 5. Conclusion

In this paper a novel approach that uses artificial neural networks and support vector machines to reduce the amount of data that is required to calculate the time domain average of a gear vibration signal is presented. Two models are proposed. Using Model 1 the data for calculating the Time Domain Average was reduced to 25 percent of the data required to calculate the Time Domain Average by direct averaging. Model 2 was found to be excellent at estimating the Time Domain Average but had the disadvantage of requiring more operations to execute. When using Model 2, less than 20 percent of the data are needed to be stored in the data logger at any given time. Furthermore, the suitability of the developed models for diagnostic purposes was assessed. It was observed that the performances of the Model 1 and Model 2 were similar over the entire life of the gear. The good performance of Model 2 can be attributed to the fact that Model 2 uses the whole data set for training and simulation, whereas Model 1 uses only a section of the data set. Using the whole data set during training and simulation exposes the formulations in the model to all the transient effects in the data, resulting in a more accurate estimate of the Time Domain Average. The performance of Model 1 relies on the generalization capabilities of the formulation used.

## 6. References

[1] L. Hongxing, Z. Hongfu, J. Chengyu and Q. Liangheng, 2000 *Mechanical Systems and Signal Processing* 14(2), pp. 279-285, An improved algorithm for direct time domain averaging.

[2] C.R. Trimble, 1968 *Hewlett-Packard Journal* 19(8), pp. 2-7. What is signal averaging?

[3] S. Braun, 1975 *Acustica* (32), pp. 69-77, The extraction of periodic waveforms by time domain averaging.

[4] S. Braun and B. Seth, 1980 *Journal of Sound and Vibration* 70(4), pp. 513-526 Analysis of repetitive mechanism signatures.

[5] P. D. McFadden, 1987 *Mechanical Systems and Signal Processing* (1), pp. 83-95, A revised model for the extraction of periodic waveforms by time domain averaging.

[6] P. D. McFadden, 1989 *Mechanical Systems and Signal Processing* (3), pp. 87-97, Interpolation techniques for time domain averaging of gear vibration.

[7] B. Samanta, 2004 *Mechanical Systems and Signal Processing* (18), pp. 625-644, Gear fault detection using artificial neural networks and support vector machines.

[8] C.M. Bishop, 1995 *Neural networks for pattern recognition*, Oxford: Clarendon Press.

[9] S.R. Gunn, 1998 *Technical report*, University of Southampton, Department of Electrical and Computer Science, UK, Support vector machines for classification and regression.

[10] S. Haykin, 1999 *Neural networks*, 2$^{nd}$ edition, New Jersey, USA: Prentice-Hall Inc.

[11] V.N. Vapnik, 1999 *IEEE Transactions on Neural Networks* 10, pp. 988-1000, An overview of learning theory.

[12] V.N. Vapnik, 1995 *The nature of statistical learning theory*, New York, USA: Springer-Verlag.

[13] C.J. Stander and P.S Heyns, 2002 Proceedings of the 15$^{th}$ International Congress on Condition Monitoring and Diagnostic Engineering Management, Birmingham UK, 2-4 September 2002, pp. 220-230. Instantaneous shaft speed monitoring of gearboxes under fluctuating load conditions.

[14] C.J. Stander and P.S Heyns, 2002 *Mechanical Systems and Signal Processing* 16(6) pp. 1005-1024. Using vibration monitoring for local fault detection on gears operating under fluctuating load conditions.

[15] A.D. Raath, 1992 *Structural dynamic response reconstruction in the time domain*, Ph.D. thesis, Department of Mechanical and Aeronautical Engineering, University of Pretoria.

[16] M. P. Norton, 1989 *Fundamentals of noise and vibration analysis for engineers*, New York: Cambridge University Press.